\documentclass[12pt]{article}
\usepackage{latexsym}
\usepackage{graphicx}
\usepackage{caption2}
\usepackage{psfrag}
\usepackage{amsmath}
\newcommand{\be}{\begin{equation}}
\newcommand{\ee}{\end{equation}}
\newcommand{\bea}{\begin{eqnarray}}
\newcommand{\eea}{\end{eqnarray}}

\newcommand{\lesssim}{ {\
\lower-1.2pt\vbox{\hbox{\rlap{$<$}\lower5pt\vbox{\hbox{$\sim$}}}}\ } }
\newcommand{\gtrsim}{ {\
\lower-1.2pt\vbox{\hbox{\rlap{$>$}\lower5pt\vbox{\hbox{$\sim$}}}}\ } }
\long\def\symbolfootnote[#1]#2{\begingroup%
\def\thefootnote{\fnsymbol{footnote}}\footnote[#1]{#2}\endgroup}

\input epsf

\begin{document}

\begin{titlepage}

\begin{flushright}
%\today
\end{flushright}
\vspace*{1.5cm}
\begin{center}
{\Large \bf The operator product expansion does not imply parity doubling of
hadrons}\\[2.0cm]

{\bf Oscar Cat\`{a}, Maarten Golterman\symbolfootnote[1]{Permanent address:
Department of Physics and Astronomy, San Francisco State University, 1600 Holloway
Ave, San Francisco, CA 94132, USA
} } and {\bf Santiago  Peris}\\[1cm]

Grup de F{\'\i}sica Te{\`o}rica and IFAE\\ Universitat
Aut{\`o}noma de Barcelona, 08193 Barcelona, Spain

\end{center}

\vspace*{1.0cm}

\begin{abstract}

We examine whether the fact that QCD is chirally invariant at short distances
necessarily leads to the prediction that hadrons form approximate parity doublets,
as has been recently put forward by Glozman and collaborators.
We show that this is not the case, and
we exhibit some of the pitfalls of trying to link the operator product expansion
to the hadron spectrum.
We illustrate our arguments within a model for
scalar and pseudo-scalar mesons used  recently by Shifman to argue for
parity doubling.  We find that, whatever the experimental situation may be,
there is no theoretical basis for parity doubling based on the use of the
operator product expansion.

\end{abstract}

\end{titlepage}

\setcounter{footnote}{0}

\section{Introduction}

In recent years there has been a renewed interest in the possibility that ``chiral
symmetry restoration" takes place for the higher excited hadronic states of a given
spin and isospin.  For each spin and isospin, resonances occur with both parities,
and if chiral symmetry, $SU(n_f)_L\times SU(n_f)_R$, would be linearly realized  these
resonances would have to organize themselves in equal-mass pairs of opposite parity.
Of course, it is well known that chiral symmetry is spontaneously broken, and that
there is therefore no symmetry reason that such parity pairing should occur.   This
observation goes back to the work of Ref.~\cite{ccwz}, and was recently re-emphasized
in the language of effective field theory in Ref.~\cite{jps}.

However, one may ask whether there could be any other dynamical reason that
approximately degenerate parity pairs occur in QCD.  Loosely speaking, the intuition
would seem to be that for highly excited resonances the physics of the vacuum (which breaks
chiral symmetry) is less important.  Such resonances would then couple less strongly
to pions, and possibly organize themselves in approximate parity pairs.

Here we examine one such dynamical scenario.\footnote{For a recent discussion of
other possibilities, see Ref.~\cite{jps2}.}  The basic observation is that, since chiral
symmetry breaking is a non-perturbative phenomenon, and QCD is asymptotically
free, chiral symmetry is restored at high energies, and, {\it mutatis mutandis}, for
highly excited resonances \cite{csrrefs}.   To our knowledge, this has been most concretely explored in Ref.~\cite{shifman} in the context of the operator product expansion (OPE).

In order to be precise, we begin with setting the stage for our discussion.  First, in order to give
an unambiguous meaning to the mass of a resonance, we will consider the limit in which
$N_c$, the number of colors, is taken to infinity.  Otherwise highly excited resonances
merge into the perturbative continuum, due to their finite widths, and their individual
identity is lost.  In contrast, in the limit of infinite $N_c$, all resonances are stable,
and spectral functions are (infinite) sums over delta functions.   Second, we need a
definition of what it means for chiral symmetry to be restored for highly excited
resonances.  We consider a tower of resonances of given spin, isospin and parity,
with masses $M_{S_n}$, and the corresponding tower with the same spin and isospin,
but opposite parity, with masses $M_{P_n}$.  For definiteness, we will consider
scalar and pseudo-scalar mesons, as was done in Ref.~\cite{shifman}.\footnote{Our
arguments extend easily to the example of vector and axial-vector mesons.}  The integer
$n$ labels the successive states in each tower.    We will now assume that for each
$S$ state with label $n$, a closest $P$ state can be identified, and for the sake of
convenience we will use the same label $n$ for this $P$ state.\footnote{In other words,
we assume that approximate parity doublets can be identified, at least for large enough
$n$.}  We now define as a measure of chiral symmetry the ``spectral overlap" $\rho(n)$
by \cite{glozman}
\be
\label{spoverlap}
\rho(n)=\frac{M_{P_n}-M_{S_n}}{M_{S_{n+1}}-M_{S_n}}\ .
\ee
The choice of denominator here deserves some comment.  Obviously, we want $\rho(n)$
to be dimensionless, but we do not want $\rho(n)$ to go to zero for large $n$ trivially.
Therefore, the denominator is taken to be the smallest possible quantity with dimension of mass
made out of the masses with ``resonance number" approximately equal to $n$.
``Chiral symmetry restoration" for a particular spin and isospin
takes place if $\rho(n)$ goes to zero for large $n$.

Another measure of ``chiral symmetry restoration" considered in the literature \cite{glozman,glozman2} is the ``chiral asymmetry"
\be
\label{cs}
\chi(n)=\frac{M_{P_n}-M_{S_n}}{M_{P_n}+M_{S_n}}\ .
\ee
This is \textit{not} a good measure of chiral symmetry restoration, however, because of the
choice of the denominator.  We know that the resonance masses will increase with $n$,
and thus this denominator will grow with $n$.  That alone will make $\chi$ go to zero
for large $n$, as long as the numerator grows less fast --- in fact, typically one would envisage
defining parity pairs such that the numerator stays finite for large $n$.    Clearly,
in order to avoid this problem, one should
measure the difference in mass between potential parity partners relative to the overall
mass density (per unit of $n$), as is done by the quantity $\rho(n)$.
We will therefore not consider $\chi(n)$ in this paper.

The claim made in Ref.~\cite{shifman} can now be expressed as follows:  The slowest pattern
of chiral symmetry restoration (for the towers of scalar and pseudo-scalar mesons) allowed
by the asymptotic behavior of the OPE is given by
\be
\label{claim}
|\rho(n)|\sim \frac{1}{n}\ ,\ \ \ \ \ n\to\infty\ .
\ee
It was assumed that the asymptotic behavior for large $n$ in both the $S$ and $P$ channels is Regge
like, $M_{S_n,P_n}^2\sim n\Lambda^2$.
If Eq.~(\ref{claim}) would be true, then clearly it would be appropriate to conclude that chiral symmetry
is restored for highly excited resonances (in this channel).  This result was inferred
from a superconvergence relation,
\be
\label{superconv}
\sum_n^\infty\delta M^2_n=0\ ,
\ee
in which $\delta M_n^2=M_{S_n}^2-M_{P_n}^2$.
Superconvergence relations of this type were
also assumed to be valid in Refs.~\cite{beane,ub}.

In this note, we show, within the context of the same assumptions (large $N_c$ and Regge
behavior), that these claims do not follow from the OPE.\footnote{We have pointed out before that superconvergence
relations such as Eq.~(\ref{superconv}) cannot be derived from the OPE \cite{gp}.  However,
in view of the recent claims which keep appearing in the literature,
it seems worthwhile revisiting this issue once more.}  To clarify this point, we will,
within the context of the same assumptions,
exhibit a model of scalar and pseudo-scalar meson resonances which is consistent with
what is known about the OPE, and for which $\rho(n)$ approaches a non-zero constant for
large $n$.

This paper is organized as follows.  We begin with reviewing the argument presented in
Ref.~\cite{shifman} in the following section, and we point out why in general it is incorrect.  We then remind the reader of the fact that
large-$N_c$ QCD in two dimensions satisfies the assumptions, but violates the claims.
It therefore is a good counter example, and our explicit calculation of $\rho(n)$ for this
model invalidates statements about this model in Ref.~\cite{glozman}.
In Sec.~4, we construct a model of scalar and pseudo-scalar resonances in four dimensions
which further
demonstrates that the claims of Ref.~\cite{shifman} do not have to hold.  In our concluding section, we comment on the situation
at finite $N_c$.

\section{Scalar and pseudo-scalar mesons}

Following Ref.~\cite{shifman}, we start with the scalar and pseudo-scalar two-point functions
\be
\label{SPfunctions}
\Pi_{S,P}(q^2)=i\int dx\;e^{iqx}\;\langle0|T\left\{J_{S,P}(x)J^\dagger_{S,P}(0)\right\}|0\rangle\ ,
\ee
with $J_S(x)=\overline{d}(x)u(x)$ and $J_P(x)=\overline{d}(x)i\gamma_5u(x)$, and
consider a once-subtracted dispersion relation for one-half the difference of these two
correlation functions:
\be
\label{disp}
\Pi_{S-P}(q^2)=\frac{q^2}{\pi}\;\int_0^{\infty}dt\;\frac{\frac{1}{2}\left({\rm Im}\;\Pi_S(t)-\Pi'_P(t)\right)}{t(t-q^2-i\epsilon)}
+\Delta\Pi(0)
-\frac{B^2f_0^2}{-q^2-i\epsilon}\ ,
\ee
where the prime indicates that we do not include the pion pole inside the integral, instead writing it as a separate term.  We work in the chiral limit, $f_0$ is the pion decay constant in that limit (approximately 87~MeV),
and $B=-\langle\overline{\psi}\psi\rangle/f_0^2$.  We note that there
is no singularity in the integrals at $t=0$ because the spectral functions $\frac{1}{\pi}{\rm Im}\;\Pi_S(t)$
and $\frac{1}{\pi}{\rm Im}\;\Pi'_P(t)$ vanish at
$t=0$.  The subtraction constant $\Delta\Pi$,
\be
\label{deltapi}
\Delta\Pi(0)\equiv\frac{1}{2}\left(\Pi_S(0)-\Pi'_P(0)\right)=16B^2L_8
\ee
is fixed by
chiral symmetry, where $L_8$ is one of the order-$p^4$
couplings in the pion effective theory \cite{L8}.

Working at $N_c=\infty$, there is an infinite tower of infinitely narrow resonances in
each channel, and we thus have that
\bea
\label{towers}
\frac{1}{\pi}\;{\rm Im}\;\Pi_S(t)&=&2\sum_n^\infty F_{S_n}^2
\delta(t-M_{S_n}^2)\ ,\\
\frac{1}{\pi}\;{\rm Im}\;\Pi'_P(t)&=&2\sum_n^\infty F_{P_n}^2
\delta(t-M_{P_n}^2)\ .\nonumber
\eea
Furthermore, we will assume Regge-like behavior
asymptotically in $n$ \cite{shifman}, {\it i.e.}
\be
\label{Regge}
M_{S_n,P_n}^2= n\Lambda^2+\dots\ ,\ \ \ \ \ n\to\infty\ ,
\ee
with $\Lambda\sim 1$~GeV, and we also require that
\be
\label{parton}
F_{S_n,P_n}^2\sim\Lambda^2 M_{S_n,P_n}^2\ ,\ \ \ \ \ n\to\infty\ ,
\ee
in order guarantee the correct parton-model behavior.\footnote{We ignore $\alpha_S$
corrections to the parton-model logarithm, and, likewise, the fact that
$\Pi_S$ and $\Pi_P$ have non-trivial anomalous dimensions.}
Substituting these expressions
into Eq.~(\ref{disp}), one obtains
\bea
\label{sminp}
\Pi_{S-P}(q^2=-Q^2)
&=&\\
&&\hspace{-2cm}\Delta\Pi(0)-\frac{B^2f_0^2}{Q^2}
-Q^2\sum_n^\infty\left(\frac{F_{S_n}^2}{M_{S_n}^2
\left(Q^2+M_{S_n}^2\right)}
-\frac{F_{P_n}^2}{M_{P_n}^2
\left(Q^2+M_{P_n}^2\right)}\right)\ ,\nonumber
\eea
which is finite,  with the masses and residues of Eqs.~(\ref{Regge})
and (\ref{parton}).

Next, we define the mass splitting in a parity pair
(for convenience we label both the $S$ and $P$ states of the parity pair with
the same resonance number $n$, as already mentioned in the previous section)
\be
\label{mdiff}
\delta M_n^2=M_{S_n}^2-M_{P_n}^2\ ,
\ee
and we assume that $\delta M_n^2$ grows less rapidly than $M_n^2\equiv\left(M_{S_n}^2+M_{P_n}^2
\right)/2$ for $n\to\infty$.  Expanding Eq.~(\ref{sminp}) in $\delta M_n^2$ then yields,
with $F_{S_n,P_n}^2=\kappa\Lambda^2 M_{S_n,P_n}^2$,
where $\kappa$ is a proportionality constant fixed by the parton-model logarithm,
\be
\label{sminpexp}
\Pi_{S-P}(q^2=-Q^2)=\Delta\Pi(0)-\frac{B^2f_0^2}{Q^2}+\kappa\Lambda^2 Q^2\sum_n^\infty
\frac{\delta M_n^2}{(Q^2+M_n^2)^2}\left(1+O\left(\frac{\delta M_n^2}{M_n^2}\right)\right)\ .
\ee
This is to be compared with the leading term for large $Q^2$ in the OPE for this
two-point function:
\be
\label{ope}
\Pi_{S-P}(q^2=-Q^2)\sim\frac{\langle\overline\psi\psi\rangle^2}{Q^4}\ ,\ \ \ \ \ Q^2\to\infty\ ,
\ee
up to logarithmic corrections.  From the comparison between Eqs.~(\ref{sminpexp}) and
(\ref{ope}), Ref.~\cite{shifman} concludes that
\be
\label{superconvergence}
\sum_n^\infty\delta M_n^2<\infty
\ee
(and in fact that it vanishes), to avoid $1/Q^2$ terms in the OPE, and, from this, that
\be
\label{alt}
\delta M_n^2\sim (-1)^n\;\frac{\Lambda^2}{n}\ ,\ \ \ \ \ n\to\infty
\ee
is the slowest possible fall off with $n$.  Equations~(\ref{alt}) and (\ref{Regge}) then lead
to Eq.~(\ref{claim}).

Both these conclusions are unfounded.  First, Eq.~(\ref{superconvergence}) assumes
that the sum over $n$ in Eq.~(\ref{sminpexp}) commutes
with the expansion in powers of $1/Q^2$, and hence that the sum $\sum_n^\infty\delta M_n^2$
converges.\footnote{We refer to Ref.~\cite{gpes} for a detailed discussion on how
the sums in Eq.~(\ref{towers}) should be regulated so as to be consistent with chiral
symmetry.  Obviously, the regulator should be taken to infinity \textit{before} taking
$Q^2$ --- a physical scale -- large \cite{gp}.}  In writing Eq.~(\ref{sminpexp})
we assumed that the sum in that equation converges, which is indeed the case if $\delta M_n^2$
increases less rapidly than $n$.  But clearly this does not imply that $\sum_n^\infty\delta M_n^2$
converges, much less that it vanishes.  With the superconvergence relation gone,
there is also no ground for the second claim, Eq.~(\ref{alt}).  Asymptotic
behavior like that of Eq.~(\ref{alt}) may of course be consistent with the absence of $1/Q^2$ terms
in the OPE, but it certainly does \textit{not} follow from it.

Second,
we also see immediately that demanding that $\sum_n^\infty\delta M_n^2$
be finite leads to
inconsistent physics, because it would follow that  $\lim_{Q^2\to\infty}\Pi_{S-P}(Q^2)$ is equal to $\Delta\Pi(0)$.
The reason this is inconsistent is that chiral symmetry at low energy predicts that this
constant does not vanish according to Eq.~(\ref{deltapi}),
while perturbation theory predicts that $\Pi_{S-P}(Q^2)$ should
vanish for $Q^2\to\infty$ ({\it cf.} Eq.~(\ref{ope})), expressing the fact that perturbation theory does not know about chiral symmetry breaking.

All these problems get resolved by the simple fact that the assumptions made about the
scalar and pseudo-scalar two-point functions (large-$N_c$, Regge behavior and parton-model
behavior) can be fully consistent with the OPE, regardless of what happens to $\sum_n\delta M_n^2$.
We will demonstrate this by explicit example in Sec.~4.

\section{Absence of parity pairs in the 't~Hooft model}

Before getting to our example, it is instructive to consider the situation in
a simple solvable model, two-dimensional QCD at $N_c=\infty$ \cite{thooft}.
The asymptotic behavior of the meson spectrum in this model is known:
at large resonance number, the masses scale like
\be
\label{2dmass}
M_n^2\sim n\Lambda^2\ ,\ \ \ \ \ n\to\infty\ ,
\ee
and resonances of opposite parity alternate: resonances with odd $n$ have the
same parity, which, in turn, is opposite to that of resonances with even $n$.

It follows that the numerator of Eq.~(\ref{spoverlap}) is given by
\be
\label{mdiff}
M_{n+1}-M_n=\frac{M_{n+1}^2-M_n^2}{M_{n+1}+M_n}\sim\frac{\Lambda}{2\sqrt{n}}\ ,\ \ \ \ \ n\to\infty\ ,
\ee
while the denominator is given by
\be
\label{denom}
M_{n+2}-M_n\sim\frac{\Lambda}{\sqrt{n}}\ , \ \ \ \ \ n\to\infty\ .
\ee
In the 't~Hooft model, therefore,
\be
\label{rho2d}
\lim_{n\to\infty}|\rho(n)|=\frac{1}{2}\ .
\ee
While mass differences decrease as a function of $n$, there is clearly no way in
which parity pairs can be unambiguously identified.  This is reflected in the result
for the spectral overlap, Eq.~(\ref{rho2d}):  the value $1/2$ is the maximum
possible value that $\rho(n)$ can have.   This means that asymptotically the pseudo-scalar states lie
precisely halfway two successive scalar states.\footnote{A
larger value would imply that parity pairs have been misidentified.}  In this
sense, chiral symmetry is maximally broken by all meson resonances in the
`t~Hooft model, unlike what is claimed in Refs.~\cite{glozman,glozman2}.

We note that the quantity $\chi(n)$, defined in Eq.~(\ref{cs}), goes like
\be
\label{chi2d}
\chi(n)\sim \frac{1}{4n}\ ,\ \ \ \ \ n\to\infty\ ,
\ee
simply because the denominator of $\chi(n)$ goes like $2\sqrt{n}\Lambda$.
The spectrum gets more crowded with higher $n$, and
thus naturally the pseudo-scalar states get more squeezed between the
scalar states.
The 't Hooft model thus
confirms that $\chi(n)$ is not a good measure of chiral symmetry restoration.

\section{A counter example in four dimensions}

Returning to four dimensions, we will now construct an example of a mass
spectrum which satisfies all OPE
constraints, but which does not satisfy Eqs.~(\ref{superconvergence}) and (\ref{alt}).  Our model will be that of
Sec.~2, with the specific choice
\bea
\label{modelmass}
M_{S_n,P_n}^2&=&m_{S,P}^2+n\Lambda^2\ ,\ \ \ \ \ n=1,\dots\ ,\\
F_{S_n,P_n}^2&=&\kappa\Lambda^2 M_{S_n,P_n}^2\ ,\ \ \ \ \ n=1,\dots\ ,\nonumber
\eea
where $\kappa$ is a numerical constant, equal to $N_c/32\pi^2$ in large-$N_c$ QCD, and we
choose $m_S\ne m_P$.
In the pseudo-scalar sector, there is an additional massless pion pole, with residue $2B^2f_0^2$
({\it cf.} Eq.~(\ref{disp})).

Using
\be
\label{sum}
\lim_{N\to\infty}\sum_{n=1}^N\left(\frac{1}{z+n}-\frac{1}{n}\right)=-\psi(z)-\frac{1}{z}-\gamma_E\ ,
\ee
in which $\gamma_E$ is the Euler--Mascheroni constant and $\psi(z)$ is the digamma
function
\be
\label{psi}
\psi(z)=\int_0^\infty dt\;\left(\frac{e^{-t}}{t}-\frac{e^{-zt}}{1-e^{-t}}\right)=\frac{d}{dz}\log\Gamma(z)\ ,
\ee
we find that
\bea
\label{sminpexpl}
\Pi_{S-P}(q^2=-Q^2)&=&\\
&&\hspace{-4cm}\Delta\Pi(0)-\frac{B^2f_0^2}{Q^2}
+\kappa Q^2\left[\psi\left(\frac{Q^2+m_S^2}{\Lambda^2}\right)
-\psi\left(\frac{Q^2+m_P^2}{\Lambda^2}\right)+\frac{\Lambda^2}{Q^2+m_S^2}
-\frac{\Lambda^2}{Q^2+m_P^2}\right]\nonumber\\
&&\hspace{-4cm}=\Delta\Pi(0)+\kappa(m_S^2-m_P^2)-\left(B^2f_0^2+\frac{1}{2}\kappa(m_S^4-m_P^4+\Lambda^2(m_S^2-m_P^2))\right)
\frac{1}{Q^2}\nonumber\\
&&\hspace{-3cm}+\kappa
\left(\frac{1}{3}(m_S^6-m_P^6)+\frac{1}{2}\Lambda^2(m_S^4-m_P^4)+\frac{1}{6}\Lambda^4(m_S^2-m_P^2)\right)\frac{1}{Q^4}+O\left(\frac{1}{Q^6}\right)\ ,\nonumber
\eea
where we used
\be
\label{psiexp}
\psi(z)=\log{z}-\frac{1}{2z}-\sum_{n=1}^\infty\frac{B_{2n}}{2nz^{2n}}\ ,
\ee
valid for Re~$z>0$ ($B_{2n}$ are the Bernoulli numbers).  Requiring that this be equal to the
leading OPE expression, $-6\pi\alpha_S\langle\overline\psi\psi\rangle^2/Q^4$, leads to the
relations
\bea
\label{operelations}
16B^2L_8&=&\kappa(m_P^2-m_S^2)\ ,\\
B^2f_0^2&=&\frac{1}{2}\kappa(m_P^4-m_S^4+\Lambda^2(m_P^2-m_S^2))\ ,\nonumber\\
6\pi\alpha_S\langle\overline\psi\psi\rangle^2&=&\frac{1}{6}\kappa
\left(2(m_P^6-m_S^6)+3\Lambda^2(m_P^4-m_S^4)+\Lambda^4(m_P^2-m_S^2)\right)\ ,\nonumber
\eea
where we used Eq.~(\ref{deltapi}).  Clearly, our model wants $m_P^2$ to be larger than $m_S^2$,
which is not inconsistent with what is known about the lowest lying resonances in the
scalar and pseudo-scalar channels.

This example demonstrates several facts.  First, in this model, $\delta M_n^2=m_S^2-m_P^2$
for all $n=1,\dots,\infty$, and therefore $\sum_{n=1}^\infty\delta M_n^2$ does not converge.
Nevertheless, there is no clash with the OPE.  Second, the coefficients
of the OPE appearing in Eq.~(\ref{operelations}) are related to low-energy constants and
know about resonances at all scales, and not just about those for
asymptotically large $n$.
Clearly, if we would include other low-lying resonances which are not part of the
equally-spaced towers, their masses and residues would show up in these sum rules as well.
Our example satisfies all the assumptions of Ref.~\cite{shifman}, but invalidates
the claims.   In particular, it is straightforward to see that this model does not show
chiral symmetry restoration:
\be
\label{modelrho}
\lim_{n\to\infty}\rho(n)={\rm min}\left\{\frac{m_P^2-m_S^2}{\Lambda^2},
\frac{m_S^2+\Lambda^2-m_P^2}{\Lambda^2}\right\}
\ee
(assuming that $m_S^2<m_P^2<m_S^2+\Lambda^2$).

\section{Discussion}

In this paper, we have restricted ourselves to $N_c=\infty$, because in that case all resonances
in a certain channel are stable, and their masses are thus well-defined.  Within this framework,
we have demonstrated that from what we know about the short-distance behavior of QCD (in this limit)
nothing can be inferred about parity doubling in the hadron spectrum.
The basic reason for the difficulty comes from the following pitfall:
even though it is true that perturbation theory
tells us that chiral symmetry breaking effects die off at high energy, this is true only
in the euclidean regime.  The spectrum lies in the Minkowski regime, and
perturbation theory clearly fails badly for infinite $N_c$, predicting a smooth
continuum where there is in fact an infinite tower of Dirac delta functions.
The lack of a straightforward connection between the two regimes is at the origin
of the difficulty.
We illustrated the absence of parity doubling
in detail in the context of a Regge-like model for scalar and pseudo-scalar mesons
which satisfies both the requirements of large $N_c$ in the Minkowski regime and
those of leading-order perturbation theory in the euclidean regime.  We
see no reason why the situation would be different in channels with different
spin and isospin.

A further fact illustrated by our example is that the coefficients of the OPE know
about \textit{all} resonances in a certain channel, and not only about the asymptotic behavior
for large $n$.  This is as one would expect from the OPE, which  can be
used to connect information about hadron spectra as a whole to the behavior of correlation
functions deep in the euclidean regime.   The fact that this connection must work
constrains the possibilities for the spectrum quite generally.  For instance, in Sec.~2 we showed
that a superconvergence relation of the form of Eq.~(\ref{superconv}) cannot hold
for large $N_c$,
because it is in conflict with perturbation theory.

In the model we considered in Sec.~4, potential candidates for parity doublers can still
be identified, because the slope in scalar and pseudo-scalar channels is taken to be
equal, as was done in Ref.~\cite{shifman}.  This leads to a definite prediction for $\rho(n)$.
However, we do not really know whether the slope in the scalar and pseudo-scalar
channels is equal, and one way to generalize the model is to allow for $\Lambda_S\ne\Lambda_P$.
Even in this case, a set of sum rules generalizing Eq.~(\ref{operelations}) can still be
derived, and thus there is no conflict with the OPE.  But in this case it clearly makes little sense to try identify potential parity partners, let alone define $\rho(n)$.

Another important generalization is to go to the world of finite $N_c$.   Assuming that at
finite but large $N_c$ the masses of the model of Sec.~4
pick up decay widths $\Gamma_n$ which scale like $\sim\sqrt{n}/N_c$ \cite{bsz},
it can be shown \cite{bsz,cgs} that  as
a consequence in this model the spectral function
$\frac{1}{\pi}\;{\rm Im}\;(\Pi_S(t)-\Pi_P(t))$ exhibits a fall off $\sim 1/t^2$, for large $t$.\footnote{For
$\Pi_{V-A}$ one finds a fall off $1/t^3$ \cite{cgs},
but a similar analysis gives a fall off $1/t^2$ for $\Pi_{S-P}$.}
It would however simply be incorrect to interpret this fall off in the spectral function as a sign
of chiral symmetry restoration between the scalar and pseudo-scalar channels.
It is just the growing of the widths with excitation number which causes this effect,
by washing out the delta functions in Eq.~(\ref{towers}), and thus increasing the
overlap between opposite-parity doublets.  The mass spectrum (defined as the
location of the real part of the masses) is still given by Eq.~(\ref{modelmass}), and
no parity alignment takes place.

While the details of this argument are based
on the finite-$N_c$ version of our model, it is generally expected that  broadening
of resonances toward larger $n$ takes place also in QCD.  This smoothens the spectral functions,
ameliorating the connection to the perturbative regime, and one expects that
the difference of the $S$ and $P$ spectral functions will fall off bounded by some power
of $t$ also in the real world.  However, our model demonstrates that this does not
imply that chiral symmetry restoration has to take place.  The slowest possible pattern of
chiral symmetry restoration compatible with the OPE is no chiral symmetry
restoration at all.

\section*{Acknowledgements}

We thank Matthias Jamin for helpful comments, and Eduardo de Rafael for reading the
manuscript.
OC and SP are supported in part by CICYT-FEDER-FPA2005-02211 and SGR2005-00916.
MG is supported in part by the Generalitat de Catalunya under the program
PIV1-2005 and by the US Department of Energy.

\end{document}